\documentclass{elsarticle}

\usepackage{hyperref}

\journal{LAA}

\usepackage{dsfont}
\usepackage{amsmath}
\usepackage{amssymb}

\usepackage{amsfonts}
\usepackage{graphicx}
\usepackage{epstopdf}
\usepackage{algorithmic}
\usepackage{caption}

\ifpdf
  \DeclareGraphicsExtensions{.eps,.pdf,.png,.jpg}
\else
  \DeclareGraphicsExtensions{.eps}
\fi

\usepackage{geometry}
\newcommand{\be}{\begin{equation}}
\newcommand{\ee}{\end{equation}}
\newcommand{\ba}{\left [ \begin{array}}
\newcommand{\ea}{\end{array} \right ]}
\newcommand{\bea}{\begin{eqnarray}}
\newcommand{\eea}{\end{eqnarray}}

\newcommand{\rank}{\mathop{\mathrm{rank}}}
\newcommand{\diag}{\mathop{\mathrm{diag}}}
\newcommand{\real}{\mathop{\mathrm{Re}}}
\newcommand{\image}{\mathop{\mathrm{Im}}}
\def\iu{\ensuremath{\mathrm{i}}}

\begin{document}

\begin{frontmatter}

\title{On Recursive Computation of Coprime Factorizations of Rational Matrices\tnoteref{mytitlenote}}
\tnotetext[mytitlenote]{Contribution to the LAA Special Issue in Honor of Paul Van Dooren}

\author{Andreas Varga}
\address{Gilching, Germany}

%
%

\begin{abstract}
General computational methods based on descriptor state-space realizations are proposed to compute coprime factorizations of rational matrices with minimum degree denominators. The new methods rely on  recursive pole dislocation techniques, which allow to successively place all poles of the factors into a ``good'' region of the complex plane. The resulting McMillan degree of the denominator factor is equal to the number of poles lying in the complementary ``bad'' region and therefore is minimal. The developed pole dislocation techniques are instrumental for devising numerically reliable procedures  for the computation of  coprime factorizations with proper and stable factors of arbitrary improper rational matrices and coprime factorizations with inner denominators. Implementation aspects of the proposed algorithms are discussed and illustrative examples are given.
\end{abstract}


\end{frontmatter}


%
%
%

\section{Introduction}
Consider a disjunct partition of the complex plane $\mathds{C}$ as
\be\label{Cgoodbad}  \mathds{C} = \mathds{C}_g \cup \mathds{C}_b, \quad \mathds{C}_g \cap \mathds{C}_b = \emptyset \, ,\ee
where we only assume $\infty \not\in \mathds{C}_g$.
Any rational matrix $G(\lambda)$ can be expressed in a left fractional form
\be\label{lf} G(\lambda) = M^{-1}(\lambda)N(\lambda) \, , \ee
or in a right fractional form
\be\label{rf} G(\lambda) = N(\lambda)M^{-1}(\lambda) \, ,\ee
where both the denominator factor $M(\lambda)$ and the numerator factor $N(\lambda)$ have only (finite) poles in $\mathds{C}_g$. Any method to compute a right factorization can be used to compute a left factorization by simply applying it to the transposed matrix $G^T(\lambda)$. Therefore, in this paper we only focus on right factorizations of the form (\ref{rf}).

Of special interest are the so-called coprime factorizations, where the factors satisfy additional conditions.  A fractional representation of $G(\lambda)$ of the form (\ref{rf}), with $N(\lambda)$ and $M(\lambda)$ having only poles in $\mathds{C}_g$, is a \emph{right coprime factorization} (RCF) over $\mathds{C}_g$, if there exist $U(\lambda)$ and $V(\lambda)$ with poles only in $\mathds{C}_g$ which satisfy
\( U(\lambda)M(\lambda)+V(\lambda)N(\lambda) = I \, .\)
Coprime factorizations with minimum-degree denominators satisfy additionally the condition that the number of poles of the resulting $M(\lambda)$ is equal to the number of poles of $G(\lambda)$ in $\mathds{C}_b$. When counting poles, both finite and infinite poles are counted with their multiplicities. When determining minimum-degree coprime factorizations over $\mathds{C}_g$, all poles of $M(\lambda)$ can be arbitrarily chosen from $\mathds{C}_g$. The coprime factorizations over a ``good'' domain of poles $\mathds{C}_g$, also including factorizations with special properties of the denominator factor (e.g., stability, innerness), are important in extending various controller synthesis methods \cite{Zhou96} to more general systems or in solving synthesis problems of fault detection filters \cite{Varg17} in the most general setting.

If $G(\lambda)$ represents the \emph{transfer-function matrix} (TFM) of a (generalized) linear time-invariant system, then the variable $\lambda$ stands either for the complex variable $s$ in the Laplace transform, in the case of a continuous-time system, or for the complex variable $z$ in the $Z$-transform, in the case of a discrete-time system. In this case, $G(\lambda)$ has real coefficients (i.e., all entries are ratios of polynomials with real coefficients) and $\mathds{C}_g$ and $\mathds{C}_b$ are chosen symmetrically located with respect to the real axis, with $\mathds{C}_g$ having at least one point on the real axis. This guarantees the existence of real factorizations, where both factors $N(\lambda)$ and $M(\lambda)$  have real coefficients.
Typical choices for $\mathds{C}_g$ are the open left half-plane of $\mathds{C}$ for a continuous-time system, or the open unit disc centered in the origin for a discrete-time system. In these cases, the corresponding factorizations are called proper and stable RCFs. An important class of such factorizations is the class of stable and proper RCF with inner denominators, where $M(\lambda)$ is additionally an inner TFM satisfying the all-pass property $M^\sim (\lambda)M(\lambda) = I$. Recall that, the conjugate TFM $M^\sim (\lambda)$ is defined in a continuous-time setting as $M^{\sim }(s) = M^T(-s)$, while in a discrete-time setting $M^\sim (z) = M^T(1/z)$. A TFM $M(\lambda)$ which satisfies only $M^\sim (\lambda)M(\lambda) = I$ is called all-pass. Therefore, an inner TFM is a stable all-pass TFM.

For the computation of right coprime factorizations with minimum-degree denominators a recursive pole-dislocation-based approach can be employed. Inspired by the ideas of Belevitch \cite{Bele68}, such a method has been developed in \cite{Vand77} and later refined in \cite{Door90}. The basic approach can be simply described as an $n_b$-step procedure, where $n_b$ is the number of poles of $G(\lambda)$ in $\mathds{C}_b$.
To compute the RCF of $G(\lambda)$, it is possible to find a sequence of $n_b$ nonsingular rational matrices $\widetilde M_i(\lambda)$, $i = 1, \ldots, n_b$, each of McMillan degree 1, with one (arbitrary) pole in $\mathds{C}_g$ and one (fixed) zero in $\mathds{C}_b$, such that the sequences $N_i(\lambda) := N_{i-1}(\lambda)\widetilde M_i(\lambda)$ and $M_i(\lambda) := M_{i-1}(\lambda)\widetilde M_i(\lambda)$ for $i = 1, \ldots, n_b$, initialized with $N_0(\lambda) = G(\lambda)$ and $M_0(\lambda) = I$, generate the factors $N(\lambda) := N_{n_b}(\lambda)$ and $M(\lambda) := M_{n_b}(\lambda)$ of the RCF (\ref{rf}). The zero of $\widetilde M_i(\lambda)$ is chosen to cancel with a pole of $N_{i-1}(\lambda)$ lying in $\mathds{C}_b$, such that after $n_b$ steps, all poles of $G(\lambda)$ are canceled and dislocated to values in $\mathds{C}_g$. This approach can be also employed when additionally imposing that all elementary factors $\widetilde M_i(\lambda)$ are inner. In the case of complex poles, the above technique leads, in general, to complex factorizations. Therefore, to obtain  real factorizations,  for each complex conjugate pair of poles a second-degree real factor can be used to cancel simultaneously both poles. Second-degree factors may also be necessary when dislocating a pair of real poles into a pair of complex conjugate poles. General formulas for constructing first- and second-degree factors are given in \cite{Vand77}. In the case of an improper $G(\lambda)$, similar first-degree formulas can be devised for canceling a single infinite pole.

In this paper, we describe general descriptor state-space realization based versions of the above recursive approach, which eliminate  the delicate computations involving  the manipulation of rational matrices. The recursive pole dislocation techniques underlying the new general algorithms  have been developed in the spirit of the approach described in \cite{Door90} and extend the methods proposed in \cite{Varg98a} by addressing the dislocation of infinite poles.  Alternative, non-iterative approaches to compute coprime factorizations with inner denominators have been proposed in \cite{Oara99b,Oara05} and involve the solution of generalized Lyapunov equations.


\section{Descriptor system based computations}

In this section we present some basic notions on descriptor systems, and describe the main reduction and updating techniques employed in the paper.

\subsection{Preliminaries on descriptor systems}
Let $G(\lambda)$ be a $p\times m$ real rational matrix  and consider an $n$-th order descriptor system realization $(A-\lambda E,B,C,D)$, with $A-\lambda E$ an $n\times n$ regular pencil (i.e., $\det(A-\lambda E) \not\equiv 0$), which satisfies
\be\label{GTFM} G(\lambda) = C(\lambda E-A)^{-1}B+D .\ee
We will also use the equivalent notation for the TFM (\ref{GTFM})
\be\label{GTFMalt} G(\lambda) = \ba{c|c} A-\lambda E & B\\ \hline C & D \ea .\ee
If $Q, Z \in \mathds{R}^{n\times n}$ are invertible matrices, then the two realizations $(A-\lambda E,B,C,D)$ and $(\widetilde A-\lambda \widetilde E,\widetilde B,\widetilde C,D)$, whose matrices are related by a similarity transformation of the form
\be\label{dssim} \widetilde A-\lambda \widetilde E = Q(A-\lambda E)Z, \quad \widetilde B = QB, \quad \widetilde C = CZ \, ,\ee
have the same TFM $G(\lambda)$.

We recall from \cite{Verg79,Verg81} some basic notions related to descriptor system realizations.  A  realization $(A-\lambda E,B,C,D)$ is \emph{minimal} if it is controllable, observable and has no non-dynamic modes. The poles of $G(\lambda)$ are related to $\Lambda(A-\lambda E)$, the eigenvalues of the pencil $A-\lambda E$ (also known as the generalized eigenvalues of the pair $(A,E)$).
For a minimal  realization, the finite poles of $G(\lambda)$ are the finite eigenvalues of $A-\lambda E$, while the multiplicities of the infinite eigenvalues of $A-\lambda E$  are in excess one to the multiplicities of infinite poles. A finite eigenvalue $\lambda_f$ is controllable if $\rank\,[\,A-\lambda_f E \; B\,] = n$, otherwise it is uncontrollable. Similarly, a finite eigenvalue $\lambda_f$ is observable if $\rank \left[\begin{smallmatrix} A-\lambda_f E \\C \end{smallmatrix}\right] = n$, otherwise it is unobservable. Infinite controllability requires that $\rank\,[\, E \; B\,] = n$, while infinite observability requires that $\rank \left[\begin{smallmatrix} E \\C \end{smallmatrix}\right] = n$. The lack of non-dynamic modes can be equivalently expressed as $A \ker(E) \subseteq \image(E)$.

Consider a given disjunct partition of the complex plane as in (\ref{Cgoodbad}). We say the descriptor system (\ref{GTFMalt}) is \emph{proper} $\mathds{C}_g$-\emph{stable} if all  finite eigenvalues of $A-\lambda E$ belong to $\mathds{C}_g$ and all infinite eigenvalues of $A-\lambda E$ are simple. The descriptor system (\ref{GTFMalt}) (or equivalently the pair $(A-\lambda E,B)$) is $\mathds{C}_b$-\emph{stabilizable} if $\rank [\, A-\lambda E \; B \,] = n$ for all finite $\lambda \in \mathds{C}_b$. The descriptor system (\ref{GTFMalt})   (or equivalently the pair $(A-\lambda E,C)$)  is $\mathds{C}_b$-\emph{detectable} if $\rank \left[\begin{smallmatrix} A-\lambda E \\C \end{smallmatrix}\right] = n$ for all finite $\lambda \in \mathds{C}_b$.

\subsection{Reduction of the pole pencil to a special form}

Using two orthogonal transformation matrices $Q$ and $Z$, it is possible to reduce a regular pencil $A-\lambda E$ to a specially ordered \emph{generalized real Schur form} (GRSF) $\widetilde A -\lambda \widetilde E := Q(A-\lambda E)Z$, with the eigenvalues of the component diagonal blocks split in four diagonal blocks:
\be\label{inf-good-bad-spec}\hspace*{-1.5mm} \widetilde A -\lambda \widetilde E = {\arraycolsep=0.2mm\ba{cccc} A_\infty & \ast &\ast&\ast \\ 0 & A_g\!-\!\lambda E_g & \ast &\ast\\ 0 & 0 & A_{f,b}\! -\! \lambda E_{f,b} &\ast \\ 0 & 0 & 0 & A_{\infty,b}\! -\! \lambda E_{\infty,b}
\ea} , \ee
where:  $(i)$ $A_{\infty}$ is an $(n-r)\times (n-r)$ invertible (upper triangular) matrix, with $r = \rank  E$; the pair $(A_\infty,0)$ contains all simple infinite eigenvalues of $A-\lambda E$ corresponding to first-order eigenvectors; $(ii)$ $A_g$ and $E_g$ are $n_g\times n_g$ matrices, such that the pair $(A_g,E_g)$ is in a GRSF (i.e., $A_g$ upper quasi-triangular and $E_g$ upper triangular)  and $\Lambda(A_g-\lambda E_g) \subset \mathds{C}_g$; $(iii)$ $A_{f,b}$ and $E_{f,b}$ are $n_b^f\times n_b^f$ matrices, such that the pair $(A_{f,b},E_{f,b})$ is in a GRSF and  $\Lambda(A_{f,b}-\lambda E_{f,b})$ contains the finite eigenvalues lying in $\mathds{C}_b$; and $(iv)$ $A_{\infty,b}$ and $E_{\infty,b}$ are $n_b^\infty\times n_b^\infty$ upper triangular matrices and $\Lambda(A_{\infty,b}-\lambda E_{\infty,b})$ contains the rest of infinite eigenvalues. The matrices $E_g$, $E_{f,b}$, $A_{\infty,b}$ are invertible, while $E_{\infty,b}$ is nilpotent.

The reduction of a pair $(A,E)$ to the specially ordered  GRSF (\ref{inf-good-bad-spec}) can be performed in three main steps. In the first step, we isolate the infinite eigenvalues having first order eigenvectors using two orthogonal transformation matrices $Q_1$ and $Z_1$ such that
\[Q_1(A-\lambda E)Z_1 = \ba{cc} A_\infty & \ast \\ 0 &   A_{22}-\lambda E_{22} \ea , \]
where the regularity of $A-\lambda E$ ensures that $A_\infty$ is nonsingular  and the pencil $A_{22}-\lambda E_{22}$ is regular. This step involves the compression of the columns of $E$ using an orthogonal transformation matrix $Z_1$ such that $E Z_1 = [\, 0 \; E_2 \,]$, with $E_2$ full column rank $r = \rank E$, and the conformably partitioned $AZ_1 = [\, A_1 \; A_2 \,]$, with $A_1$ having full column rank $n-r$. Typically, such a column compression of $E$ is performed using the rank-revealing QR decomposition with column pivoting applied to $E^T$ \cite{Golu13}.  Then, an orthogonal matrix $Q_1$ is determined such that
$Q_1A_1 = \left[\begin{smallmatrix} A_{\infty} \\ 0 \end{smallmatrix}\right]$,
with $A_{\infty}$ an $(n-r)\times (n-r)$ invertible upper triangular  matrix. This computation can be done using the standard QR decomposition of a full  column rank matrix \cite{Golu13}.

At the second step, the pair $(A_{22}, E_{22})$ is reduced, using two orthogonal transformation matrices $Q_2$ and $Z_2$, to the form
\be\label{finite-inf} Q_2(A_{22}-\!\lambda E_{22})Z_2 =  {\ba{cc} A_{f}\!-\!\lambda E_{f} & \ast \\ 0 & A_{\infty,b}\! -\! \lambda E_{\infty,b}
\ea} , \ee
where the pair $(A_{\infty,b},E_{\infty,b})$ has only infinite generalized eigenvalues, $A_{\infty,b}\! -\! \lambda E_{\infty,b}$ is upper triangular, with $E_{\infty,b}$ nilpotent, and all eigenvalues of $A_{f}\!-\!\lambda E_{f}$ are finite. The infinite-finite splitting of eigenvalues can be reliably computed by applying the orthogonal reduction algorithm  of \cite{Oara97} to the transposed pencil $A_{22}^T-\lambda E^T_{22}$. This algorithm determines the orthogonal transformation matrices $Z_2^T$ and $Q_2^T$, such that the reduced pencil $Z_{2}^T(A_{22}^T-\lambda E^T_{22})Q_2^T$  is in a $2\times 2$ block upper triangular form, where the leading block contains the infinite eigenvalues, with the matrices in an upper triangular (staircase) form, while the trailing block contains the finite eigenvalues. The form (\ref{finite-inf}) is obtained by pertransposing the resulting pair,\footnote{Pertransposing a square matrix $M$ is the operation to form $PM^T\!P$, where $P$ is  a permutation matrix with ones down the secondary diagonal.} using a permutation matrix $P_2$ of appropriate size. This operation interchanges the order of the infinite and finite blocks and preserves the upper triangular shape of the blocks containing the infinite eigenvalues, which form the resulting pencil $A_{\infty,b}\! -\! \lambda E_{\infty,b}$.

At the third step, we reduce the pair $(A_{f}, E_{f})$
to an ordered GRSF, using two orthogonal transformation matrices $Q_3$ and $Z_3$, such that
\[ Q_3(A_{f}-\!\lambda E_{f})Z_3 =  {\ba{cc} A_g\!-\!\lambda E_g & \ast \\  0 & A_{f,b}\! -\! \lambda E_{f,b}
\ea} ,\]
where $\Lambda(A_g-\lambda E_g) \subset \mathds{C}_g$ and $\Lambda(A_{f,b}-\lambda E_{f,b}) \subset \mathds{C}_b$.
Suitable algorithms (e.g., the QZ algorithm) to reduce a matrix pair to the GRSF and to reorder the eigenvalues by permuting the $1\times1$ and $2\times 2$ diagonal blocks are described in \cite{Golu13}.

The following procedure summarizes the main computational steps of the reduction of a regular pair $(A,E)$, to a pair $(\widetilde A,\widetilde E) = (QAZ,QEZ)$ in the specially ordered GRSF (\ref{inf-good-bad-spec}), by employing the orthogonal transformation matrices $Q$ and $Z$.\\

\vspace*{2mm}
\hrule
\vspace*{2mm}
\noindent\textbf{Procedure GSORSF.}
\vspace*{2mm}
\hrule
\vspace*{2mm}
\begin{algorithmic}[1]
\item[1.] Compute an orthogonal $Z_1$ such that $E Z_1 = [\, 0 \; E_2\,]$, with $E_2$ having full column rank $r = \rank E$; compute the conformably partitioned $AZ_1 = [\, A_1 \; A_2\,]$, with $A_1$ having full column rank $n-r$.
\item[2.] Compute an orthogonal $Q_1$ such that $Q_1A_1 = \left[\begin{smallmatrix} A_{\infty} \\ 0 \end{smallmatrix}\right]$, with $A_{\infty}$ an $(n-r)\times (n-r)$ invertible upper triangular  matrix; compute the conformably partitioned matrices\\[-2mm]
    \[ Q_1A_2  = \ba{c} \ast\\ A_{22} \ea, \quad Q_1E_2  = \ba{c} \ast\\ E_{22} \ea \, .\]
\item[3.] Apply the staircase algorithm of \cite{Oara97} to the transposed pencil $A_{22}^T-\lambda E_{22}^T$ to obtain the orthogonal transformation matrices $Z_2^T$ and $Q_2^T$ such that  \\[-2mm]
\[ P_2Q_2(A_{22}-\lambda E_{22})Z_2P_2 =
{\ba{cc} A_{f}\!-\!\lambda E_{f} & \ast \\ 0 & A_{\infty,b}\! -\! \lambda E_{\infty,b}
\ea} , \]
where $P_2$ is a permutation matrix of appropriate size with ones down on the secondary diagonal, the pair $(A_{\infty,b},E_{\infty,b})$ has only infinite generalized eigenvalues, $A_{\infty,b}\! -\! \lambda E_{\infty,b}$ is upper triangular, with $E_{\infty,b}$ nilpotent, and all eigenvalues of $A_{f}\!-\!\lambda E_{f}$ are finite.
\item[4.] Compute orthogonal $Q_3$ and $Z_3$ such that\\[-2mm]
\[ Q_3(A_{f}-\!\lambda E_{f})Z_3 =  {\ba{cc} A_g\!-\!\lambda E_g & \ast \\  0 & A_{f,b}\! -\! \lambda E_{f,b}
\ea} \]
is in a GRSF, where $\Lambda(A_g-\lambda E_g) \subset \mathds{C}_g$ and $\Lambda(A_{f,b}-\lambda E_{f,b}) \subset \mathds{C}_b$.
\item[5.] Set $Q = \diag(I_{n-r},Q_3,I)\diag(I_{n-r},P_2Q_2)Q_1$, $Z = Z_1 \diag(I_{n-r},Z_2P_2)\diag(I_{n-r},Z_3,I)$ and define $\widetilde A = QAZ$ and $\widetilde E = QEZ$ from the reduced pencil (\ref{inf-good-bad-spec}).
\end{algorithmic}
\vspace*{2mm}
\hrule
\vspace*{2mm}

\vspace*{2mm}
The computations performed at Steps 1, 2 and 4 of \textbf{Procedure GSORSF} rely on standard numerically stable algorithms (described, for example, in  \cite{Golu13}), as the rank revealing QR decomposition with column pivoting or the QZ algorithm to compute and reorder the GRSF of a matrix pair. The numerical complexity of these steps is $\mathcal{O}(n^3)$. The finite-infinite separation performed at Step 3, using the numerically stable staircase algorithm of \cite{Oara97}, employs a rank revealing QR-decomposition-based staircase reduction technique. The performed orthogonal transformations consist of sequences of Householder or Givens transformations, which are not accumulated but directly applied to the involved submatrices. Using this reduction technique (i.e., without explicitly accumulating the performed orthogonal transformations employed for rank determinations), guarantees  the $\mathcal{O}(n^3)$ computational complexity of this step. Therefore, the overall computational complexity of the \textbf{Procedure GSORSF} is $\mathcal{O}(n^3)$ as well.
The reliability of rank decisions can be improved, using \emph{singular value decomposition} (SVD)-based rank decisions instead QR-decomposition-based rank decisions. Such an approach has been proposed, for example, in \cite{Door79a}. Due to the need to explicitly accumulate the performed orthogonal transformations, the worst-case complexity of this approach is $\mathcal{O}(n^4)$. An alternative computational technique of complexity $\mathcal{O}(n^3)$, with the same level of reliability of rank determinations, has been suggested in \cite{Oara97}, and consists in combining QR decomposition based reductions (without column pivoting) with SVD-based rank decisions (performed only for small matrices).


\subsection{Descriptor system based updating formulas}

The denominator factor $M(\lambda)$ of the RCF
can be represented in a product form as
\be\label{M-factored} M(\lambda) = \widetilde M_1(\lambda)\widetilde M_2(\lambda) \cdots \widetilde M_{k}(\lambda) , \ee
where each $\widetilde M_i(\lambda)$ ($i = 1, \ldots, k$) is an invertible elementary proper factor having either a real pole or a pair of complex conjugate poles.
The recursive computational procedure can be formalized as $k$ successive applications of the updating formula
\be\label{NiMi-upd} \ba{c} N_i(\lambda)\\ M_i(\lambda) \ea  = \ba{c} N_{i-1}(\lambda)\\  M_{i-1}(\lambda)\ea \widetilde M_i(\lambda) ,\quad i = 1, \ldots, k \, ,\ee
initialized with $N_0(\lambda) = G(\lambda)$ and $M_0(\lambda) = I_m$. Then, $N(\lambda) = N_k(\lambda)$ and $M(\lambda) = M_k(\lambda)$. By this approach, it is automatically achieved that the resulting $M(\lambda)$ has the least achievable McMillan degree $n_b$.

We can derive state-space formulas for the efficient implementation of the updating operations in (\ref{NiMi-upd}).
Assume $N_{i-1}(\lambda)$ and $M_{i-1}(\lambda)$ have the descriptor realizations
\be\label{Ni-1Mi-1} \ba{c} N_{i-1}(\lambda) \\ M_{i-1}(\lambda) \ea = \ba{cc|c} A_{11}-\lambda E_{11} & A_{12} -\lambda E_{12} & B_1 \\  0 & A_{22}-\lambda E_{22} & B_2 \\ \hline C_{N,1} & C_{N,2} & D_N \\ C_{M,1} & C_{M,2} & D_M  \ea =: \ba{c|c} \widetilde A-\lambda \widetilde E & \widetilde B \\ \hline \\[-3.5mm]\widetilde C_N & \widetilde D_N \\ \widetilde C_M & \widetilde D_M \ea , \ee
where $\Lambda(A_{22}-\lambda E_{22}) \subset \mathds{C}_b$. We assume that $A_{22}-\lambda E_{22}$ is a $1\times 1$ pencil in the case when $A_{22}-\lambda E_{22}$ has a real or an infinite eigenvalue, or is a $2\times 2$ pencil, in the case when $A_{22}-\lambda E_{22}$ has a pair of complex conjugate eigenvalues. This form automatically results if the pair $(\widetilde A,\widetilde E)$ is in the specially ordered GRSF (\ref{inf-good-bad-spec}). We discuss several cases which can be encountered when performing the updating.

If $B_2 = 0$, then the eigenvalue(s) of $A_{22}-\lambda E_{22}$ is (are) not controllable, and thus can be removed to obtain lower order realizations $N_{i}(\lambda) := N_{i-1}(\lambda)$ and $M_{i}(\lambda) := M_{i-1}(\lambda)$
\be\label{Ni-1Mi-1red} \hspace*{-1mm}{\arraycolsep = 0.5mm\ba{c} N_{i}(\lambda) \\ M_{i}(\lambda) \ea = \ba{c|c} A_{11}\!-\!\lambda E_{11}  & B_1 \\  \hline C_{N,1}  & D_N \\ C_{M,1} &  D_M  \ea
=: \ba{c|c} \widetilde A\!-\!\lambda \widetilde E & \widetilde B \\ \hline \\[-3.5mm]\widetilde C_N & \widetilde D_N \\ \widetilde C_M & \widetilde D_M \ea } . \ee

If $B_2 \not = 0$ and  the pencil $A_{22}-\lambda E_{22}$ has finite eigenvalues (i.e., $E_{22}$ is invertible), then the pair $( A_{22}-\lambda E_{22}, B_2)$ is (finite) controllable and there exists $F_2$ such that the eigenvalues of $A_{22}+B_2F_2-\lambda E_{22}$ can be placed in arbitrary locations in $\mathds{C}_g$. Assume that such an $F_2$ has been determined and define the elementary factor
\be\label{tildeM-fin} \widetilde M_i(\lambda) = (A_{22}+B_2F_2-\lambda E_{22},\,B_2W,\, F_2,\, W), \ee
where $W$ is chosen to ensure the invertibility of $\widetilde M_i(\lambda)$. To compute stable and proper RCFs, the choice $W = I_m$ is always possible. However, alternative choices of $W$ are necessary to ensure, for example, that $\widetilde M_i(\lambda)$ is inner (see Section \ref{sec:rcfid}). It is easy to check that the updated factors $N_i(\lambda)$  and $M_i(\lambda)$ in (\ref{NiMi-upd}) have the realizations
\[ {\arraycolsep=0.6mm\ba{c} N_i(\lambda)\\ M_i(\lambda) \ea
= \ba{cc|c} A_{11}-\lambda E_{11} & A_{12}+B_1F_2 -\lambda E_{12} & B_1W \\  0 & A_{22}+B_2F_2-\lambda E_{22} & B_2W \\ \hline C_{N,1} & C_{N,2}+ D_NF_2 & D_NW \\ C_{M,1} & C_{M,2}+D_MF_2 & D_MW  \ea } . \]
If we denote $\widetilde F = [\, 0 \; F_2\,]$, then the above relations lead to the following updating formulas \cite{Varg98a}:
\be\label{GRCF_upd1} \begin{array}{lcl} \widetilde A & \leftarrow & \widetilde A+\widetilde B\widetilde F , \\
\widetilde B &\leftarrow & \widetilde BW ,\\
\widetilde C_N & \leftarrow & \widetilde C_N+\widetilde D_N\widetilde F , \\
\widetilde C_M & \leftarrow & \widetilde C_M+\widetilde D_M\widetilde F ,\\
\widetilde D_N &\leftarrow & \widetilde D_NW ,\\
\widetilde D_M &\leftarrow & \widetilde D_MW .\end{array} \ee

If $B_2 \not = 0$ and  the $1\times 1$  pencil $A_{22}-\lambda E_{22}$ has an infinite eigenvalue (i.e., $E_{22} = 0$), then we choose the elementary factor
\be\label{Mtilde-inf} \widetilde M_i(\lambda) = (\gamma-\lambda \eta,\,B_2,\, F_2,\, W), \ee
where $\eta \not= 0$ and $\gamma/\eta$ is an arbitrary real pole in $\mathds{C}_g$, $W$ is a projection matrix chosen such $B_2W = 0$ and $\rank \left[\begin{smallmatrix} B_2\\ W \end{smallmatrix}\right] = m$, and $F_2$ is chosen such that $B_2F_2 = -A_{22}$ and $\rank [\,F_2\; W \,] = m$ (the rank conditions guarantee the invertibility of $\widetilde M_i(\lambda)$). Straightforward  choices of $F_2$ and $W$ are, for example,
\be\label{eq:f2-w-disc-inf} \begin{array}{l}
   F_2 = -B_2^T(B_2^{}B_2^T)^{-1}A_{22}^{} ,\\
   W = I-B_2^{T}(B_2^{}B_2^T)^{-1}B_2^{} .
      \end{array} \ee
By this choice of $\widetilde M_i(\lambda)$, we made the infinite eigenvalue in the realization of the updated factors $N_i(\lambda)$ and $M_i(\lambda)$ uncontrollable, and after its elimination, we obtain the realizations
\[ {\arraycolsep=0.5mm\ba{c} N_i(\lambda)\\ M_i(\lambda) \ea
= \ba{cc|c} A_{11}-\lambda E_{11} & A_{12}+B_1F_2 -\lambda E_{12} & B_1W \\  0 & \gamma-\lambda \eta& B_2 \\ \hline C_{N,1} & C_{N,2}+ D_NF_2 & D_NW \\ C_{M,1} & C_{M,2}+D_MF_2 & D_MW  \ea } .\]
The above relations lead to the following updating formulas:
\be\label{GRCF_upd2} \begin{array}{lcl}
\widetilde A & \leftarrow & \left[\begin{smallmatrix} A_{11} & A_{12}+B_1F_2 \\0 & \gamma \end{smallmatrix}\right] ,  \\ \\[-3.5mm]
\widetilde E & \leftarrow & \left[\begin{smallmatrix} E_{11} & E_{12} \\0 & \eta \end{smallmatrix}\right] , \\ \\[-3.5mm]
\widetilde B & \leftarrow & \left[\begin{smallmatrix}  B_1W \\ B_2 \end{smallmatrix}\right] , \\
\widetilde C_N & \leftarrow & \big[\, C_{N,1} \;\; C_{N,2} + D_NF_2 \,\big] ,\\
\widetilde C_M & \leftarrow & \big[\,  C_{M,1} \;\; C_{M,2} + D_MF_2  \,\big] ,\\
\widetilde D_N &\leftarrow &  D_NW ,\\
\widetilde D_M &\leftarrow &  D_MW .\end{array}  \ee

The updating techniques relying on the formulas (\ref{GRCF_upd1}) and (\ref{GRCF_upd2}) ensure that, if the original pair $(\widetilde A,\widetilde E)$ was in a GRSF,   then the updated pair will have a similar form, possibly with $\widetilde A-\lambda \widetilde E$ having a $2\times 2$ trailing block which corresponds to two real generalized  eigenvalues (to recover the GRSF, such a block can be further split into two $1\times 1$ blocks using an orthogonal similarity transformation).

\section{Computation of RCFs}
In this section we present the computational procedures to compute proper $\mathds{C}_g$-stable RCFs and stable RCFs with inner denominators, and discuss the main numerical features of the proposed algorithms, as generality, numerical reliability and computational efficiency.

\subsection{Computation of proper $\mathds{C}_g$-stable RCFs}

To compute a proper $\mathds{C}_g$-stable RCF of an arbitrary rational TFM  $G(\lambda)$ with a descriptor system realization $(A-\lambda E,B,C,D)$, we start by reducing the pair $(A,E)$ to the special GRSF (\ref{inf-good-bad-spec}) by performing a system similarity transformation as in (\ref{dssim}) using orthogonal transformation matrices $Q$ and $Z$. Then, by employing the updating techniques described previously, we can dislocate the generalized eigenvalues of the trailing elementary $1\times 1$ or $2\times 2$ blocks  to locations in $\mathds{C}_g$. Finally, the reordering of the diagonal blocks in the GRSF of the updated pair $(\widetilde A,\widetilde E)$, brings in the trailing position a new block, whose generalized eigenvalues (finite or infinite) lie in $\mathds{C}_b$. This eigenvalue dislocation process is repeated until all eigenvalues are moved into $\mathds{C}_g$.


The following procedure computes for an arbitrary $p\times m$ rational TFM  $G(\lambda)$, with a descriptor system realization $(A-\lambda E,B,C,D)$, and for a given disjunct partition of the complex plane $\mathds{C}$ as in (\ref{Cgoodbad}), a RCF  $G(\lambda) = N(\lambda) M^{-1}(\lambda)$, with  the resulting factors having proper $\mathds{C}_g$-stable   descriptor realizations $N(\lambda) =
(\widetilde{A}-\lambda\widetilde{E},\widetilde{B},\widetilde C_N,\widetilde D_N)$ and $ M(\lambda) = (\widetilde{A}-\lambda\widetilde{E},\widetilde{B},\widetilde C_M,\widetilde D_M)$. \\[5mm]

\vspace*{2mm}
\hrule
\vspace*{2mm}
\noindent\textbf{Procedure GRCF.}
\vspace*{2mm}
\hrule
\vspace*{2mm}
\begin{algorithmic}[1]
      \item[1.] Compute, using \textbf{Procedure GSORSF}, the orthogonal matrices $Q$ and $Z$ to reduce   the pair $(A,E)$   to the specially ordered GRSF $(\widetilde A,\widetilde E)$ in (\ref{inf-good-bad-spec}).
        Compute
           $\widetilde B := QB$, $\widetilde C_N := CZ$, and set $\widetilde C_M = 0$, $\widetilde D_N = D$, $\widetilde D_M = I_m$, $q := r+n_g$, $n_b = n_b^f+n_b^\infty$.
      \item[2.] If $n_b = 0$, \textbf{Exit}.
      \item[3.] Let $(A_{22},E_{22})$  be the last $k\times k$
         diagonal blocks of the GRSF of $(\widetilde A,\widetilde E)$ (with $k\! = \!1$ or $k\! =\! 2$) and let $B_2$ be the $k \times m$ matrix
         formed from the last $k$ rows of $\widetilde B$. \\ If $B_2 = 0$, then
         remove the uncontrollable eigenvalues $\Lambda(A_{22}-\lambda E_{22})$ and redefine $\widetilde A$, $\widetilde E$, $\widetilde B$, $\widetilde C_N$, $\widetilde C_M$ according to (\ref{Ni-1Mi-1red}); update $n \leftarrow n-k$, $n_b \leftarrow n_b-k$  and go to Step 2.
      \item[4.] If $E_{22} = 0$,  compute $F_2 = -B_2^T(B_2^{}B_2^T)^{-1}A_{22}^{}$ and $W = I_m-B_2^T(B_2^{}B_2^T)^{-1}B_2^{}$. Choose $\eta = 1$ and
         $\gamma \in \mathds{C}_g$ and update $\widetilde A$, $\widetilde E$, $\widetilde B$, $\widetilde C_N$, $\widetilde D_N$, $\widetilde C_M$ and $\widetilde D_M$ using  (\ref{GRCF_upd2}).
      \item[5.] If $E_{22} \not= 0$, determine $F_2$ such that $\Lambda(A_{22}+B_2F_2-\lambda E_{22}) \subset \mathds{C}_g$.
         Set $\widetilde F = [\, 0 \;\; F_2 \, ]$, $W = I_m$ and update $\widetilde A$, $\widetilde E$, $\widetilde B$, $\widetilde C_N$, $\widetilde D_N$, $\widetilde C_M$ and $\widetilde D_M$ using  (\ref{GRCF_upd1}).
%
      \item[6.] Compute the orthogonal matrices
$\widetilde{Q}$ and $\widetilde{Z}$ to  move  the last  blocks  of
$(\widetilde A,\widetilde E)$ to  positions $(q+1,q+1)$ by interchanging
the diagonal blocks of the GRSF. Compute
         $\widetilde{A} \leftarrow \widetilde{Q}\widetilde{A}\widetilde{Z}$,
         $\widetilde{E} \leftarrow \widetilde{Q}\widetilde{E}\widetilde{Z}$,
         $\widetilde{B} \leftarrow \widetilde{Q}\widetilde{B}$,
         $\widetilde{C}_N \leftarrow \widetilde{C}_N\widetilde{Z}$,
         $\widetilde{C}_M \leftarrow \widetilde{C}_M\widetilde{Z}$.
     Put $q \leftarrow q + k$, $n_b \leftarrow n_b-k$  and go to Step 2.
\end{algorithmic}
\vspace*{2mm}
\hrule
\vspace*{2mm}

With the special GRSF (\ref{inf-good-bad-spec}) computed at Step 1, the procedure executes at the beginning the Steps 3, 4 and 6 repeatedly, until all controllable infinite eigenvalues are dislocated to finite locations. In the reordering of diagonal blocks performed at Step 6 to recover the special GRSF (\ref{inf-good-bad-spec}), it is possible to exploit the structure of the blocks containing the infinite generalized eigenvalues of the pair $(\widetilde A,\widetilde E)$. Consequently, the swapping of two $1\times 1$ diagonal blocks, where the first one contains an infinite and the second one contains a finite eigenvalue, can be reliably performed by explicitly restoring the zero diagonal elements in $\widetilde E$ (these elements are usually blurred by roundoff errors incurred by swapping). This part of the algorithm ensures that the resulting factors are proper and can be interpreted as a more elaborated version of both of the Algorithms PRRCF1 and PRRCF2 presented in \cite{Varg98a}.

After the completion of the dislocation of infinite eigenvalues, the procedure executes the Steps 3, 5 and 6 repeatedly, until all controllable finite eigenvalues in $\mathds{C}_b$ are dislocated to new locations in $\mathds{C}_g$. This computation corresponds to Algorithm GRCF-P in \cite{Varg98a} and  can be interpreted as an extension of the GRSF-based method for pole assignment \cite{Varg95}.

The \textbf{Procedure GRCF}  is completely general, being applicable regardless the original descriptor realization of $G(\lambda)$ is $\mathds{C}_b$-stabilizable or not, is infinite controllable or not. The resulting pair $(\widetilde A, \widetilde E)$ is in a particular GRSF with $n-r$ simple infinite eigenvalues in the leading $n-r$ positions (no such block exists if $E$ is invertible). Thus, the elimination of non-dynamic (simple) modes can be done by applying standard state residualization formulas.

A minimal realization of the  denominator $M(\lambda)$
can be easily determined.
The resulting $\widetilde C_M$ has always the form
\be \widetilde C_M = [ \, 0 \;\; \widetilde C_{M,2} \, ],\label{CM-min}\ee
where the number of columns of $\widetilde C_{M,2}$ is equal to the number of
controllable generalized eigenvalues of the pair $(A,E)$ lying in $\mathds{C}_b$. By
partitioning accordingly the resulting $\widetilde {E}$, $\widetilde {A}$ and $\widetilde {B}$
\be {\arraycolsep=.45mm
           \widetilde {A} = \ba{cc}
                {A}_{11}&{A}_{12} \\
                0     &{A}_{22}
                \ea ,\;
            \widetilde {E} = \ba{cc}
                {E}_{11} & {E}_{12} \\
                0      & {E}_{22}
               \ea ,\;
           \widetilde {B} = \ba{c}
                {B}_{1} \\
                {B}_{2}
                \ea , } \label{eab-min}
\ee
we obtain $({A}_{22}-\lambda {E}_{22},{B}_{2},\widetilde C_{M,2},\widetilde D_{M})$, which
is a minimal descriptor system realization of $M(\lambda)$. The order
of the minimal realization of $M(\lambda)$ has the least
possible McMillan degree if all
eigenvalues of $A-\lambda E$ in $\mathds{C}_b$ are observable (i.e, the pair $(A-\lambda E,C)$ is $\mathds{C}_b$-detectable).

\subsection{Computation of stable RCFs with inner denominator} \label{sec:rcfid}
For the computation of stable RCFs with inner denominators, a recursive procedure, similar to that of previous section, can be developed.
In this case, we use the partition of the complex plane with $\mathds{C}_g = \mathds{C}_s$ and $\mathds{C}_b = \mathds{C}\setminus \mathds{C}_s$, where $\mathds{C}_s$ is the stability domain for the poles. Recall, that $\mathds{C}_s$ is the open left half-plane for a continuous-time system, or the interior of the unit circle centered in the origin, for a discrete-time system.  For a given rational matrix $G(\lambda)$, a necessary and sufficient condition for the existence of a stable RCF with inner denominator is that $G(\lambda)$ has no poles in  $\partial \mathds{C}_s$, the boundary of $\mathds{C}_s$. In the continuous-time case, this means that $G(s)$ is proper and has no poles on the imaginary axis, while in the discrete-time case $G(z)$ has no poles on the unit circle centered in the origin, but for the sake of generality, $G(z)$ can be improper (i.e., having infinite poles).

For a descriptor system realization $(A-\lambda E,B,C,D)$ of $G(\lambda)$, sufficient conditions for the existence of a stable RCF with inner denominator can be formulated in terms of the eigenvalues of the pole pencil $A-\lambda E$. In the continuous-time case, we merely require that the pencil $A-s E$ has no finite  controllable eigenvalues on the imaginary axis and no higher order controllable infinite eigenvalues. In the discrete-time case, the pencil  $A-z E$ has no controllable eigenvalues on the unit circle centered in the origin, but may have arbitrary number of infinite eigenvalues.

In the recursive procedure to compute the RCF with inner denominator, elementary inner factors are employed to dislocate one or a pair of complex conjugate poles at each iteration. Let us assume that at the $i$-th step, we have the matrices $A_{22}$, $E_{22}$ and $B_2$ as defined in (\ref{Ni-1Mi-1}). We encounter two cases when determining the $i$-th elementary factor $\widetilde M_i(\lambda)$ in (\ref{M-factored}).

If $B_2 \not = 0$ and $E_{22}$ invertible, then we choose $\widetilde M_i(\lambda)$ of the form (\ref{tildeM-fin}), with $F_2$ and $W$ determined as follows (see \cite[Fact 5]{Varg98a}): for a continuous-time system
\be\label{eq:f2-w-cont}    F_2 = -B_2^{T}(YE_{22}^T)^{-1} ,\quad W = I_m ,
 \ee
where $Y$ is the solution of the \emph{generalized continuous-time Lyapunov equation} (GCLE)
\be\label{eq:y-cont}   A_{22}^{}YE_{22}^T + E_{22}^{}YA_{22}^{T} -B_2^{}B_2^{T} = 0 ,  \ee
while for a discrete-time system $F_2$ and $W$ are computed from
\be\label{eq:f2-w-disc}  \begin{array}{l}
   F_2 = -B_2^T(YA_{22}^T)^{-1} ,\\
   W^T\big(I+B_2^{T}(E_{22}^{}YE_{22}^T)^{-1}B_2^{}\big)W = I ,
      \end{array} \ee
where $Y$ is the solution of the \emph{generalized discrete-time Lyapunov  equation} (GDLE) 
\be\label{eq:y-disc}
   A_{22}^{}YA_{22}^{T} -B_2^{}B_2^{T} = E_{22}^{}YE_{22}^T .
\ee
The above choice ensures that the poles of $\widetilde M_i(\lambda)$ are the reflected generalized eigenvalues of the pair $(A_{22},E_{22})$ with respect to the imaginary axis, in the continuous-time case, or with respect to the unit circle in the origin, in the discrete-time case. For this case, the updating formulas (\ref{GRCF_upd1}) are used.

The second case may appear only for an improper discrete-time system, for which $B_2 \not = 0$, $A_{22} \not = 0$, and $E_{22} = 0$. The corresponding infinite pole is reflected to the origin, by choosing  $\widetilde M_i(z)$ of the form (\ref{Mtilde-inf}), with $\gamma = 0$, $\eta = -A_{22}$, and $F_2$ and $W$ determined as in (\ref{eq:f2-w-disc-inf}).
For this case, the updating formulas (\ref{GRCF_upd2}) are used. The formulas (\ref{eq:f2-w-disc-inf}) underlying the computation of the elementary inner factor $\widetilde M_i(z)$ have been derived using the dual results of \cite[Theorem 3.2]{Oara05}.


The following procedure computes for a $p\times m$ rational TFM  $G(\lambda)$, with a descriptor system realization $(A-\lambda E,B,C,D)$, a RCF  with inner denominator $G(\lambda) = N(\lambda) M^{-1}(\lambda)$, with  the resulting factors having proper $\mathds{C}_s$-stable   descriptor realizations $N(\lambda) =
(\widetilde{A}-\lambda\widetilde{E},\widetilde{B},\widetilde C_N,\widetilde D_N)$ and $ M(\lambda) = (\widetilde{A}-\lambda\widetilde{E},\widetilde{B},\widetilde C_M,\widetilde D_M)$.

\vspace*{2mm}
\hspace*{-1.43em}\begin{minipage}{\textwidth}
\hrule
\vspace*{2mm}
\noindent\textbf{Procedure GRCFID.}
\vspace*{2mm}
\hrule
\vspace*{2mm}

\begin{algorithmic}[1]
\item[1.] Compute, using \textbf{Procedure GSORSF}, the orthogonal matrices $Q$ and $Z$ to reduce   the pair $(A,E)$   to the specially ordered GRSF $(\widetilde A,\widetilde E)$ in (\ref{inf-good-bad-spec}).
        Compute
           $\widetilde B := QB$, $\widetilde C_N := CZ$, and set $\widetilde C_M = 0$, $\widetilde D_N = D$, $\widetilde D_M = I_m$, $q := r+n_g$, $n_b = n_b^f+n_b^\infty$.
\item[2.] If $n_b = 0$, \textbf{Exit}.
\item[3.] Let $(A_{22},E_{22})$  be the last $k\times k$
         diagonal blocks of the GRSF of $(\widetilde A,\widetilde E)$ (with $k\! = \!1$ or $k\! =\! 2$) and let $B_2$ be the $k \times m$ matrix
         formed from the last $k$ rows of $\widetilde B$. \\ If $B_2 = 0$, then
         remove the uncontrollable eigenvalues $\Lambda(A_{22}-\lambda E_{22})$ and redefine $\widetilde A$, $\widetilde E$, $\widetilde B$, $\widetilde C_N$, $\widetilde C_M$ according to (\ref{Ni-1Mi-1red}); update $n \leftarrow n-k$, $n_b \leftarrow n_b-k$  and go to Step 2.
      \item[4.] If $E_{22} = 0$, then  if the descriptor system is continuous-time, \textbf{Exit} (no solution exists); else, set
         $\gamma =0$, $\eta = -A_{22}$,  compute $F_2$ and $W$ according to (\ref{eq:f2-w-disc-inf}),  and update $\widetilde A$, $\widetilde E$, $\widetilde B$, $\widetilde C_N$, $\widetilde D_N$, $\widetilde C_M$ and $\widetilde D_M$ using  (\ref{GRCF_upd2}).
      \item[5.] If $E_{22} \not= 0$,  then  if $\Lambda(A_{22}-\lambda E_{22}) \subset \partial\mathds{C}_s$, \textbf{Exit} (no solution exists); else, determine $F_2$ and $W$ as in (\ref{eq:f2-w-cont}) for a continuous-time system, or as in (\ref{eq:f2-w-disc}) for  a discrete-time system, set $\widetilde F = [\, 0 \; F_2 \, ]$,  and update $\widetilde A$, $\widetilde E$, $\widetilde B$, $\widetilde C_N$, $\widetilde D_N$, $\widetilde C_M$ and $\widetilde D_M$ using  (\ref{GRCF_upd1}).
%
      \item[6.] Compute the orthogonal matrices
$\widetilde{Q}$ and $\widetilde{Z}$ to  move  the last  blocks  of
$(\widetilde A,\widetilde E)$ to  positions $(q+1,q+1)$ by interchanging
the diagonal blocks of the GRSF. Compute
         $\widetilde{A}  \leftarrow \widetilde{Q}\widetilde{A}\widetilde{Z}$,
         $\widetilde{E}  \leftarrow  \widetilde{Q}\widetilde{E}\widetilde{Z}$,
         $\widetilde{B}  \leftarrow  \widetilde{Q}\widetilde{B}$,
         $\widetilde{C}_N  \leftarrow  \widetilde{C}_N\widetilde{Z}$,
         $\widetilde{C}_M  \leftarrow  \widetilde{C}_M\widetilde{Z}$.
     Put $q  \leftarrow  q + k$, $n_b  \leftarrow  n_b-k$  and go to Step 2.
\end{algorithmic}%
\vspace*{2mm}
\hrule
\end{minipage}
\vspace*{2.5mm}

The \textbf{Procedure GRCFID}  is completely general and is able to compute a stable RCF with inner denominator, provided the existence conditions formulated in terms of the poles of $G(\lambda)$ are fulfilled. Note that uncontrollable eigenvalues in $\mathds{C}_b$ or at infinity are automatically deflated. In contrast, the alternative, non-iterative approaches proposed in \cite{Oara99b,Oara05},  are only applicable to stabilizable descriptor realizations.    The resulting pair $(\widetilde A, \widetilde E)$ is in a particular GRSF with $n-r$ simple infinite eigenvalues in the leading $n-r$ positions, which allows the immediate elimination of non-dynamic (simple) modes.
A minimal realization of the  denominator $M(\lambda)$
can be similarly determined as in the case of \textbf{Procedure GRCF} and has the least McMillan degree, provided the descriptor realization is $\mathds{C}_b$-detectable.

\emph{Remark 1.} In the particular case, when $G(\lambda)$ is a square all-pass and anti-stable TFM (i.e., with all poles in $\mathds{C}\setminus\mathds{C}_s$ and all zeros in $\mathds{C}_s$),  a RCF with inner denominator $G(\lambda) = N(\lambda)M^{-1}(\lambda)$ has the factors $N(\lambda) = I$ and $M(\lambda) =  G^\sim(\lambda)$. It follows, that all eigenvalues in the resulting realization of $N(\lambda)$ computed by the \textbf{Procedure GRCFID} will be unobservable. More generally, if the given TFM $G(\lambda)$ contains an unstable all-pass factor with, say $k$ unstable poles, then the resulting descriptor system realization of $N(\lambda)$, computed by the \textbf{Procedure GRCFID}, will contain $k$ unobservable eigenvalues, which are precisely the stable zeros of the all-pass factor. See \textbf{Example 2} presented in Section \ref{sec:examples}, with a discrete-time TFM $G(z)$ containing an unstable all-pass factor.
\subsection{Numerical aspects}
As already mentioned, the \textbf{Procedures GRCF} and \textbf{GRCFID} are completely general, being able to compute RCFs of rational matrices, independently of the properties of the underlying descriptor system realization. For the computation of RCFs with minimum degree denominator factors, the $\mathds{C}_b$-detectability condition is sufficient to guarantee the least achievable McMillan degrees of the denominators. The right coprimeness   of the computed factors follows from the full (column) rank property of $\left[\begin{smallmatrix} N(\lambda) \\ M(\lambda) \end{smallmatrix}\right]$ for all $\lambda \in \mathds{C}_g$ \cite{Vidy11}.

The \textbf{Procedure {GRCF}} is essentially a recursive pole assignment algorithm which exploits and maintains the GRSF of the pair $(\widetilde A,\widetilde E)$. It represents a specialization of the generalized pole assignment
      algorithm of \cite{Varg95} for descriptor systems or the pole assignment algorithm of \cite{Varg81b} for  standard systems (i.e., $E = I$). These algorithms are generally considered as satisfactory computational methods, as long as the partial feedback gains $F_2$ computed at Steps 4 or 5, have moderate sizes, as---for example, $\| F_2\| \leq \kappa\| A\| / \| B\|$, with say $\kappa < 100$. A careful implementation of the computation of $F_2$ in conjunction with optimal choices of the poles to be assigned can significantly contribute to the reduction of the size of $F_2$.
      Unfortunately, the above restriction on the partial feedback gains cannot be always fulfilled if
      large gains are necessary to move all poles into $\mathds{C}_g$. This may occur
      either if the  poles in the ``bad'' region $\mathds{C}_b$ are too far from those in the ``good'' region $\mathds{C}_g$
      or if these poles are weekly controllable (e.g., small value of $\|B_2\|$). Nevertheless, the \textbf{Procedure {GRCF}} can be still considered a numerically reliable algorithm, since the above norm condition on $F_2$ can be easily checked at each computational step and, therefore, the potential loss of numerical reliability can be easily detected.

The numerical properties of \textbf{Procedure {GRCFID}} are similar to those of \textbf{Procedure {GRCF}}, as long as the partial feedback gains $\|F_2\|$ at Steps 4 and 5 are reasonably small.
However, this condition for numerical reliability may not always be fulfilled due to the lack of any freedom in assigning the poles. Recall that the unstable poles are reflected into symmetrical positions with respect to $\partial\mathds{C}_s$, and this may occasionally require  large gains.

The main computation burden in the proposed algorithms is the computation of the specially ordered GRSF in (\ref{inf-good-bad-spec}) at Step 1 and the preservation of this form using eigenvalue reordering techniques at Step 6. With the use of standard algorithms (i.e., the QR decomposition based on the Householder method, the QZ algorithm to compute and reorder the GRSF), see \cite{Golu13}, and of the staircase algorithm of \cite{Oara97}, the overall numerical complexity is $\mathcal{O}(n^3)$.

\section{Implementation issues}
\label{sec:implementation}
In this section we discuss several implementation issues, which are instrumental for an efficient and robust implementation of the proposed computational procedures. The discussed issues are related to the choice of the poles to be assigned at each iteration in the \textbf{Procedure GRCF}, the use of enhanced accuracy (square-root) techniques to compute the elementary inner factors in \textbf{Procedure GRCFID}, the elimination of the non-dynamic modes of the resulting factors and the enforcement of the upper quasi-triangular shape of the resulting system matrices when computing left coprime factorizations.

\subsection{Pole selection and assignment}\label{sec:polsel}
The selection of appropriate poles to be assigned at each iteration can significantly influence the overall numerical behavior of the \textbf{Procedure GRCF}. The strategies to select poles depend on the concrete definition of the ``good'' region   $\mathds{C}_g$ (and also of its complement $\mathds{C}_b = \mathds{C}\setminus \mathds{C}_g$), and target either a stabilization oriented factorization or a pole allocation oriented factorization.

\subsubsection{Stabilization oriented factorization} A relevant measure to characterize the desired dynamics of the factors is the stability degree, say  $\alpha$, of the poles of the resulting factors. For a continuous-time system, $\alpha < 0$ and $\alpha$ represents the maximum admissible real part of poles. Consequently,  $\mathds{C}_g$  can be defined as
\be\label{Cg-cont} \mathds{C}_g = \{ s \in \mathds{C} \mid \real(s) \leq \alpha \} . \ee
For a discrete-time system, $0 \leq \alpha < 1$ and $\alpha$ represents the maximum admissible magnitude of poles. Therefore,  $\mathds{C}_g$  can be defined as
\be\label{Cg-disc} \mathds{C}_g = \{ z \in \mathds{C} \mid |z| \leq \alpha  \} . \ee
For a given stability degree $\alpha$, the assignment of poles of the elementary factors at Steps 4 and 5  of \textbf{Procedure GRCF} can be done in such a way, that partial feedback gains $F_2$ with reduced size result.

In the case of an $1\times 1$ block  $A_{22}-\lambda E_{22}$ with a finite real eigenvalue, the pole to be assigned can always be chosen $\gamma =\alpha$. This choice leads at Step 5 to a least norm of $F_2$, which can be computed by solving
\[ B_2F_2 = E_{22}\gamma -A_{22} \]
for the least-norm solution $F_2$. If $B_2$ has the RQ decomposition
\be\label{b2_RQ} B_2 = [\, \sigma \; 0 \,]V^T = \sigma V_1^T, \ee
where $\sigma \not= 0$ and $V = [\, V_1\;V_2\,]$ is orthogonal, with $V_1$ an $m\times 1$ matrix, then $F_2$ can be explicitly computed as
\be\label{eq:f2-rq} F_2 =  V_1\frac{E_{22}\gamma -A_{22}}{\sigma} . \ee

In the case of an $1\times 1$ block  $A_{22}-\lambda E_{22}$ with an infinite eigenvalue (i.e., $E_{22} = 0$), we can use the RQ decomposition (\ref{b2_RQ}) to determine the feedback gain $F_2$ and the projection matrix $W$ in (\ref{eq:f2-w-disc-inf}) as
\be\label{eq:f2-w-disc-inf-qr} \begin{array}{l}
   F_2 = -V_1\displaystyle\frac{A_{22}}{\sigma}  ,\\[3mm]
   W = I_m-V_1^{}V_1^{T} .
      \end{array} \ee

In the case of a $2\times 2$ block corresponding to a pair of complex conjugate eigenvalues $\mu \pm \iu\tau$, to reduce the size of the partial feedback $F_2$ it is desirable to perform a minimum shifting of these poles to their new locations. This can be achieved by choosing the poles to be assigned $\{\gamma_1,\gamma_2\}$ as $\alpha \pm \iu\tau$ in the continuous-time case and $\alpha(\mu \pm \iu\tau)/\sqrt{\mu^2+\tau^2}$ in the discrete-time case. The computation of $F_2$ is done depending on the rank of $B_2$.

If $\rank B_2 = 2$ and $\Theta$ is $2\times 2$ matrix such that the eigenvalues of $E_{22}^{-1}\Theta$ are the eigenvalues to be assigned, then
$F_2$ can be computed by solving
\[ B_2F_2 = E_{22}\Theta-A_{22} \]
for the least-norm solution $F_2$. This solution can be computed using the SVD of $B_2$
\be\label{eq:B2-svd} B_{2}^{} = U [\, \Sigma \; 0\,]V^T , \ee
where $\Sigma$ is a $2\times 2$ invertible diagonal matrix, and $U$ and $V$ are orthogonal matrices. If we partition $V$ as $V = [\, V_1\;V_2\,]$, with $V_1$ an $m\times 2$ matrix, then $F_2$ results as
\be\label{F2-B2-svd} F_2 = V_1\Sigma^{-1}U^T(E_{22}\Theta -A_{22}). \ee


The choice of $\Theta$ is not unique and an optimal choice would be one for which the (Frobenius) norm of $F_2$ is minimum. For a pair of complex conjugate eigenvalues $ \{\gamma_1,\gamma_2\}$, the matrix $\Theta$ can be parameterized in terms of two parameters $\theta_1$ and $\theta_2$ as
\[ \Theta(\theta_1,\theta_2) = \ba{cc} \theta_{1} & \theta_{2} \\ \displaystyle\frac{\theta_1(\gamma_1+\gamma_2-\theta_1)-\gamma_1\gamma_2}{\theta_2} & \gamma_1+\gamma_2-\theta_1 \ea .\]
If we denote with $F_2(\theta_1,\theta_2)$ the expression (\ref{F2-B2-svd}) evaluated for the above $\Theta(\theta_1,\theta_2)$, then the optimal values of the free parameters $\theta_1$ and $\theta_2$ can be determined by minimizing $\|F_2(\theta_1,\theta_2)\|_F^2$.

If $\rank B_2 = 1$, then Procedure A of \cite{Varg81b} can be applied to the standard controllable pair $(E_{22}^{-1}A_{22}^{},E_{22}^{-1}B_{2}^{})$ to determine $F_2$ which assigns the selected pair of complex conjugate poles. In this case, we compute the SVD of $E_{22}^{-1}B_{2}^{}$ as
\[ E_{22}^{-1}B_{2}^{} = U [\, \Sigma \; 0\,]V^T , \]
where $\Sigma = \ba{c} \sigma \\0\ea$, is a $2\times 1$ matrix with $\sigma  \not=0$, and $U$ and $V$ are orthogonal matrices. We compute
\be\label{eq:alpha12} U^TE_{22}^{-1}A_{22}^{}U = \ba{cc} \widetilde \alpha_{11} & \widetilde \alpha_{12} \\ \widetilde \alpha_{21} & \widetilde \alpha_{22} \ea,  \ee
where $\widetilde \alpha_{21} \not=0$ is guaranteed by the controllability of the pair $(A_{22}-\lambda E_{22},B_2)$. If we partition $V$ as $V = [\, V_1\;V_2\,]$, with $V_1$ an $m\times 1$ matrix, then the partial feedback $F_2$ is obtained as
\be\label{F2-B2-rank1}  F_2 = V\ba{c} \widetilde F \\ 0 \ea U^T = V_1\widetilde F U^T,
 \ee
where $\widetilde F = [\ \varphi_1\; \varphi_2\,]$, with $\varphi_1$ and $\varphi_2$ computed as
\[ \begin{aligned} \varphi_1 &= (\gamma_1+\gamma_2- \widetilde \alpha_{11}-\widetilde \alpha_{22})/\sigma ,\\
\varphi_2 &= (\widetilde \alpha_{22}/\widetilde \alpha_{21})\varphi_1+(\widetilde \alpha_{11}\widetilde \alpha_{22}-\widetilde \alpha_{12}\widetilde \alpha_{21}-\gamma_1\gamma_2)/(\widetilde \alpha_{21}\sigma) . \end{aligned}
\]

\subsubsection{Pole assignment oriented factorization} Let $\alpha$ be a desired stability degree for the poles of the resulting factors and define the ``good'' region $\mathds{C}_g$  as in (\ref{Cg-cont}) or (\ref{Cg-disc}), depending on the system type, continuous- or discrete-time, respectively. Simultaneously, let $\Gamma$ be a symmetric set of complex values such that $\Gamma \subset \mathds{C}_s$, which contains the desired poles  to be assigned for the two factors.
In this setting, the initial reduction of the pair $(A,E)$ to the special GRSF (\ref{inf-good-bad-spec}) ensures, that all eigenvalues of $A-\lambda E$ lying in $\mathds{C}_g$ are preserved in the descriptor system realizations of the resulting factors and only the eigenvalues lying in the ``bad'' region $\mathds{C}_b$
or at infinity are assigned to the values specified in $\Gamma$. We will assume that the number of specified poles in $\Gamma$ is greater than the sum of the number of controllable (higher order) infinite eigenvalues and  the number of controllable finite eigenvalues in $\mathds{C}_b$. Otherwise, $\Gamma$ is extended ``on the fly''  with poles chosen from $\mathds{C}_g$, in accordance with the strategy used for the stabilization oriented factorization. In determining the partial feedback $F_2$ at Steps 4 and 5  of \textbf{Procedure GRCF}, we encounter several cases according to the size of the last block $A_{22}-\lambda E_{22}$ and the chosen poles to be assigned. After each successful partial pole assignment, $\Gamma$ is updated by removing the already assigned poles.

If $A_{22}-\lambda E_{22}$ is an $1\times 1$ block with a finite eigenvalue, then a real pole $\gamma \in \Gamma$ is selected, which is the nearest one to the eigenvalue of  $A_{22}-\lambda E_{22}$. This pole is then assigned, using (\ref{b2_RQ}) and (\ref{eq:f2-rq}). In the case, when there is no real pole available in $\Gamma$, but $n_b > 1$, then a $2\times 2$ block can be formed, either by adjoining an adjacent $1\times 1$ block or by interchanging the last two blocks of the GRSF to bring a $2\times 2$ block in the last position. Then, a pair of complex conjugate poles from $\Gamma$ can be assigned (see below). If $n_b = 1$, then $\gamma = \alpha$ is assigned.

If $A_{22}-\lambda E_{22}$ has an infinite eigenvalue (i.e., $E_{22} = 0$), then $\gamma$ is selected as the nearest real pole in $\Gamma$ to the boundary $\partial\mathds{C}_s$ and the formulas for $F_2$ and $W$ in (\ref{eq:f2-w-disc-inf-qr}) are used. If no real pole is available in $\Gamma$, then $\gamma = \alpha$ is chosen. It is possible to use more complicated schemes, as for example, adjoining two adjacent blocks with infinite eigenvalues, or a finite $1\times 1$ block and an infinite block, and assigning a pair of complex conjugate poles using suitable updating formulas (still to be developed). The chosen strategy to assign only real poles for all infinite poles appears to be the simplest one, and therefore, well suited for implementation purposes. A trivial workaround to assign a pair of complex conjugate poles to replace two infinite poles, is to assign first two real poles for the two last $1\times 1$ blocks, and then to update the resulting $2\times 2$ block by assigning a pair of complex conjugate poles (see below).

If $A_{22}-\lambda E_{22}$ is a $2\times 2$ block, then we choose the poles to be assigned $\{\gamma_1,\gamma_2\}$, as the nearest ones in $\Gamma$ to the eigenvalues of $A_{22}-\lambda E_{22}$.
If $A_{22}-\lambda E_{22}$ has complex conjugate eigenvalues, then we assign either two real poles or, preferably, a pair of complex conjugate poles, while if $A_{22}-\lambda E_{22}$ resulted by adjoining two $1\times 1$ diagonal blocks, then a pair of complex conjugate poles is assigned. Depending on the rank of the corresponding $B_2$, we employ for the computation of $F_2$ either the formula (\ref{F2-B2-svd}) if $\rank B_2 = 2$ or the formula (\ref{F2-B2-rank1}) if $\rank B_2 = 1$.

If $A_{22}-\lambda E_{22}$ resulted by adjoining two $1\times 1$ diagonal blocks and $\rank B_2 = 1$, it may happen that the resulting pair $(A_{22}-\lambda E_{22},B_2)$ is not controllable (because the real eigenvalue in the leading $1\times 1$ diagonal block of $A_{22}-\lambda E_{22}$ may be uncontrollable). The lack of controllability can be easily detected, because $\widetilde \alpha_{21} = 0$ in (\ref{eq:alpha12}). The uncontrollable eigenvalue $\widetilde \alpha_{22}$ can be eliminated by performing an orthogonal similarity transformation on the pair $(A_{22}-\lambda E_{22},B_2)$ to obtain
\[ U(A_{22}-\lambda E_{22})V := \ba{cc} \alpha_{11}-\lambda \eta_{11} & \alpha_{12}-\lambda \eta_{12} \\ 0 & \alpha_{22}-\lambda \eta_{22} \ea , \quad UB_2 := \ba{c} \widetilde B_2 \\ 0 \ea , \]
with $\widetilde B_2 \not = 0$. Here, $U$ is the orthogonal matrix from the SVD of $B_2$ in (\ref{eq:B2-svd}), while $V$ is the orthogonal matrix from the RQ decomposition of $UE_{22}$ (i.e., $UE_{22}V$ is upper triangular). After the separation of the uncontrollable $1\times 1$ block $\alpha_{22}-\lambda \eta_{22}$, the updating formulas (\ref{Ni-1Mi-1red}) are used to remove this block.

\subsection{Square-root based computation of elementary inner factors} \label{sec:squareroot} In the \textbf{Procedure GRCFID}, the computation of each elementary inner factor  $\widetilde M_i(\lambda)$ as in (\ref{tildeM-fin}) involves the solution of either of the GCLE (\ref{eq:y-cont}) or of the GDLE (\ref{eq:y-disc}), from which the partial feedback $F_2$ and feedthrough matrix $W$ are determined. In each case, the solution $Y$ is positive definite, thus it can be expressed as $Y = SS^T$, where $S$ is an upper triangular matrix, also called (improperly) the \emph{square-root} of $Y$. Methods to solve Lyapunov equations directly for the square-root factor of the solution have been proposed in \cite{Hamm82} for $E_{22} = I$  and in \cite{Penz98a} for a general $E_{22}$. These methods are provably more accurate than methods which computes $Y$, because they avoid explicitly forming the matrix product $B_2^{}B_2^{T}$. These methods are applicable provided $\Lambda(A_{22}-\lambda E_{22}) \subset \mathds{C}_s$ and $B_2^{}B_2^{T}$ stays instead $-B_2^{}B_2^{T}$ in (\ref{eq:y-cont}) or  (\ref{eq:y-disc}). To cope with these restrictions, instead of solving (\ref{eq:y-cont}), we can solve
\[ -A_{22}^{}YE_{22}^T -E_{22}^{}YA_{22}^{T} +B_2^{}B_2^{T} = 0,  \]
where $\Lambda(-A_{22}-\lambda E_{22}) \subset \mathds{C}_s$, and instead solving (\ref{eq:y-disc}), we can solve
\[    E_{22}^{}YE_{22}^T - A_{22}^{}YA_{22}^{T} + B_2^{}B_2^{T} = 0, \]
where $\Lambda(E_{22}-\lambda A_{22}) \subset \mathds{C}_s$. Thus, both above Lyapunov equations can be solved directly for the square-root factor $S$.

The upper triangular shape of $S$ and the diagonal form of $E_{22}$ can be exploited in evaluating the partial feedback $F_2$ in (\ref{eq:f2-w-cont}) as
\[ F_2 = -B_2^{T}(YE_{22}^T)^{-1} = -(S^{-T}S^{-1}(E_{22}^{-1}B_2))^T . \]
Moreover, for an $1\times 1$ block, we can use that $A_{22}+B_2F_2 = -A_{22}$.
Similarly, $F_2$ in (\ref{eq:f2-w-disc}) can be computed as
\[ F_2 = -B_2^{T}(YA_{22}^T)^{-1} = -(S^{-T}S^{-1}(A_{22}^{-1}B_2))^T  \]
and for an $1\times 1$ block, we can use that $A_{22}+B_2F_2 = E_{22}^2/A_{22}^{}$.
The feedthrough term $W$ in (\ref{eq:f2-w-disc}) can be computed as $W = R^{-1}$, with $R$ an upper triangular (Cholesky) factor satisfying
\[ R^TR = I_m+B_2^{T}(E_{22}^{}YE_{22}^T)^{-1}B_2^{} = I_m + X^TX = [\, I_m\; X^T\,]\ba{c}I_m\\ X\ea ,\]
where $X =  S^{-1}E_{22}^{-1}B_2$ is a $k\times m$ matrix, with $k = 1$ or $k = 2$. To compute $R$, the initial Cholesky factor $I_m$ can be updated, by performing, $k$ times rank-1 changes, thus avoiding to form explicitly the product $X^TX$ (see \cite{Golu13}[Section 6.5.1] for suitable QR factorization based updating techniques).

\subsection{Removing non-dynamic modes} At the end of both \textbf{Procedure GRCF} and \textbf{Procedure GRCFID}, the matrices of the resulting realizations of the factors can be partitioned in the form
\be\label{final_realization} \ba{c} N(\lambda)\\ M(\lambda) \ea = {\def\arraystretch{1.4}\ba{c|c} \widetilde A-\lambda \widetilde E & \widetilde B \\ \hline
\widetilde C_{N}  & \widetilde D_{N} \\[-1mm]
\widetilde C_{M} & \widetilde D_{M} \ea}
:=
{\def\arraystretch{1.4}\ba{cc|c} A_\infty & \widetilde A_{12}-\lambda \widetilde E_{12} & \widetilde B_1 \\[-1mm]
0 & \widetilde A_{22}-\lambda \widetilde E_{22} & \widetilde B_2 \\ \hline
\widetilde C_{N,1} & \widetilde C_{N,2} & \widetilde D_{N} \\[-1mm]
0 & \widetilde C_{M,2} & \widetilde D_{M} \ea} , \ee
where $A_\infty$ is a $(n-r)\times(n-r)$ upper triangular matrix, with $r = \rank E$, and the pair $(A_\infty,0)$ contains the $n-r$ simple infinite eigenvalues (i.e., the non-dynamic modes), while the pair $(\widetilde A_{22},\widetilde E_{22})$ is in a GRSF with all eigenvalues in $\mathds{C}_g$. Using state residualization formulas, the $n-r$ simple infinite eigenvalues can be eliminated to obtain a reduced order descriptor system realization of the form
\[ \ba{c} N(\lambda)\\ M(\lambda) \ea = \ba{c|c} \widetilde A_{22}-\lambda \widetilde E_{22} & \widetilde B_2 \\ \hline \\[-3.5mm]
 \widehat C_{N} & \widehat D_{N} \\
 \widehat C_{M} & \widetilde D_{M} \ea , \]
where
\[ \begin{aligned}
\widehat C_{N} &= \widetilde C_{N,2}^{}-\widetilde C_{N,1}^{}A_\infty^{-1}\big(\widetilde A_{12}^{}-\widetilde E_{12}^{}\widetilde E_{22}^{-1}\widetilde A_{22}^{}\big) ,\\
\widehat D_{N} &= \widetilde D_N^{}-\widetilde C_{N,1}^{}A_\infty^{-1}\big(\widetilde B_1^{}-\widetilde E_{12}^{}\widetilde E_{22}^{-1}\widetilde B_2^{}\big) ,\\
\widehat C_{M} &= \widetilde C_{M,2} .
\end{aligned}
\]
\subsection{Computation of left coprime factorizations} If we apply \textbf{Procedure GRCF} or \textbf{Procedure GRCFID} to the dual realization $(A^T-\lambda E^T,C^T,B^T,D^T)$ of $G^T(\lambda)$, we obtain
the RCF  $G^T(\lambda) = N(\lambda) M^{-1}(\lambda)$, from which a left coprime factorization of $G(\lambda)$ results as $G(\lambda) = \big(M^{T}(\lambda)\big)^{-1}N^T(\lambda)$. The realizations of $N(\lambda)$ and $M(\lambda)$ have the form (\ref{final_realization}) with the pair $(\widetilde A,\widetilde E)$ in an upper GRSF. This form is advantageous in many applications. For example, the eigenvalues of $\widetilde A-\lambda\widetilde E$ can be determined practically at no computational cost, and can be further reordered by preserving the upper quasi-triangular shape (GRSF) of the pair $(\widetilde A,\widetilde E)$. The dual realizations of $N^{T}(\lambda)$ and $M^T(\lambda)$ have the pair $\big(\widetilde A^T,\widetilde E^T\big)$ in a lower GRSF, which has some inconveniences for numerical computations. For example, the computation of eigenvalues using standard tools (e.g., MATLAB\footnote{MATLAB$^\circledR$ is a registered trademark of The Mathworks, Inc.}) involves performing the whole QZ algorithm \cite{Golu13} to obtain an upper GRSF. A simple trick can be used to obtain the dual realizations with an upper GRSF. This comes down to form the realizations of $N^{T}(\lambda)$ and $M^T(\lambda)$ as
\[ [\, N^{T}(\lambda)\;M^T(\lambda)\,] =
{\def\arraystretch{1.4}\ba{c|cc} P\widetilde A^TP-\lambda P\widetilde E^TP & P\widetilde C_{N}^T & P\widetilde C_{M}^T \\ \hline \widetilde B^TP & \widetilde D_{N}^T & \widetilde D_{M}^T \ea}  ,\]
where $P$ is the permutation matrix of appropriate size with ones down on the secondary diagonal. Note that, in the above realization, the simple infinite eigenvalues are now located in the trailing position of the pair $\big(P\widetilde A^TP, P\widetilde E^TP\big)$.

\section{Numerical examples}\label{sec:examples} The proposed factorization procedures to compute RCFs of rational matrices
have been implemented as MATLAB functions and belong to the free software collection of \emph{Descriptor Systems Tools} (DSTOOLS) \cite{Varg17a}. To illustrates the capabilities of the proposed computational algorithms, we present two simple examples computed with the functions \texttt{\bfseries grcf} and \texttt{\bfseries grcfid} available in DSTOOLS, which implements the \textbf{Procedure GRCF} and \textbf{Procedure GRCFID}, respectively. Both functions call the function \texttt{\bfseries gsorsf} to compute the specially ordered GRSF. This function implements the \textbf{Procedure GSORSF}, but employs at Step 3, instead the method of \cite{Oara97}, the Algorithm 3.2.1 of \cite{Beel88}, for which an implementation is available in the SLICOT library \cite{Benn99}.

\subsection*{Example 1} Consider the continuous-time improper TFM
\[ G(s) = \left[\begin{array}{cc} s^2 & \displaystyle\frac{s}{s +1}\\[2mm]0 & \displaystyle\frac{1}{s} \end{array}\right],
\]
with the corresponding minimal realization given by
\[ \ba{c|c} A-s E & B \\ \hline C & D \ea := \left[\begin{array}{ccccc|cc}  1 & - s & 0 & 0 & 0 & 0 & 0\\ 0 & 1 & - s & 0 & 0 & 0 & 0\\ 0 & 0 & 1 & 0 & 0 & -1 & 0\\ 0 & 0 & 0 &  - 1 - s & 0 & 0 & 1\\ 0 & 0 & 0 & 0 & - s & 0 & 1\\  \hline
1 & 0 & 0 & -1 & 0 & 0 & 1\\ 0 & 0 & 0 & 0 & 1 & 0 & 0
\end{array}\right] .
 \]
$G(s)$ has the following set of poles: $\{-1, 0, \infty, \infty\}$. To compute a stable and proper RCF of $G(s)$, we employed the pole assignment oriented factorization, with a stability degree of $\alpha = -1$ and the desired set of poles $\Gamma = \{ -1, -2, -3\}$. With the option to eliminate the non-dynamic modes, the function \texttt{\bfseries grcf} computes the following  descriptor system realization of the factors (converted automatically to symbolic expressions)
\[ \ba{c}N(s)\\M(s)\ea = {\def\arraystretch{1.4}
\left[\begin{array}{cccc|cc}   -1 - s & 0 & 0 & -3 & 0 & -1\\ 0 &  -1 - s & 1 - s & 0 & 0 & 0\\
0 & 0 &  - 2 - s & 0 & \frac{\sqrt{2}}{2} & 0\\ 0 & 0 & 0 &  - 3 - s & 0 & -1\\ \hline
1 & -\frac{\sqrt{2}}{2} & \frac{5\, \sqrt{2}}{2} & 3 & -1 & 1\\ 0 & 0 & 0 & -1 & 0 & 0\\ 0 & -\frac{\sqrt{2}}{2} & -\frac{\sqrt{2}}{2} & 0 & 0 & 0\\ 0 & 0 & 0 & 3 & 0 & 1 \end{array}\right] , }
\]
which correspond to the rational matrices
\[ N(s) = \ba{cc} -\displaystyle\frac{s^2}{(s+1)(s+2)} & \displaystyle\frac{s^2}{(s+1)(s+3)} \\[3mm] 0 & \displaystyle\frac{1}{s+3} \ea , \quad
M(s) = \ba{cc} -\displaystyle\frac{1}{(s+1)(s+2)} & 0 \\[2mm] 0 & \displaystyle\frac{s}{s+3} \ea  .
\]
The McMillan degree of $M(s)$ is three, thus the least possible one.
\subsection*{Example 2} Consider the discrete-time improper TFM
\[ G(z) = \left[\begin{array}{cc} z^2 & \displaystyle\frac{z}{z - 2}\\[2mm] 0 & \displaystyle\frac{1}{z} \end{array}\right] , \]
with the corresponding minimal realization given by
\[ \ba{c|c} A-z E & B \\ \hline C & D \ea := \left[\begin{array}{ccccc|cc}
 1 & - z & 0 & 0 & 0 & 0 & 0\\ 0 & 1 & - z & 0 & 0 & 0 & 0\\ 0 & 0 & 1 & 0 & 0 & -1 & 0\\ 0 & 0 & 0 & 2 - z & 0 & 0 & 2\\ 0 & 0 & 0 & 0 & - z & 0 & 1\\ \hline
 1 & 0 & 0 & 1 & 0 & 0 & 1\\ 0 & 0 & 0 & 0 & 1 & 0 & 0 \end{array}\right]
 .
\]
$G(z)$ has the following set of poles: $\{2, 0, \infty, \infty\}$ and therefore, a RCF with inner denominator exists. With the option to eliminate the non-dynamic modes, the function \texttt{\bfseries grcfid} computes the following  descriptor system realization of the factors (converted automatically to symbolic expressions)
\[ \ba{c}N(z)\\M(z)\ea = {\def\arraystretch{1.4}
\left[\begin{array}{cccc|cc} - z & 0 & 0 & \frac{3}{4} & 0 & \frac{1}{2}\\ 0 & - \sqrt{2}\, z & \sqrt{2} & 0 & 0 & 0\\ 0 & 0 & -\frac{\sqrt{2}\, z}{2} & 0 & \frac{\sqrt{2}}{2} & 0\\ 0 & 0 & 0 & \frac{1}{2} - z & 0 & -1\\ \hline
0 & 0 & 0 & - \frac{1}{4} & 1 & \frac{1}{2}\\ 1 & 0 & 0 & 0 & 0 & 0\\ 0 & 1 & 0 & 0 & 0 & 0\\ 0 & 0 & 0 & \frac{3}{4} & 0 & \frac{1}{2} \end{array}\right] } ,\]
which correspond to the rational matrices
\[ N(z) = {\ba{cc} 1 & \displaystyle\frac{1}{2z-1} \\[3mm] 0 & \displaystyle\frac{z-2}{z(2z-1)} \ea } , \quad
M(z) = \ba{cc} \displaystyle\frac{1}{z^2} &  0 \\[1mm] 0 & \displaystyle\frac{z-2}{2z-1}\ea .
\]
The McMillan degree of $M(z)$ is three, thus the least possible one. Interestingly, the McMillan degree of $N(z)$ is only two, because two unobservable eigenvalues in 0 have been removed. These eigenvalues are the zeros of the (improper) all-pass factor $\diag(z^2,1)$ with two infinite poles, which is contained in $G(z)$.

\section{Conclusions}
\label{sec:conclusions}

In this paper we proposed two numerically reliable algorithms to compute right coprime factorizations of rational matrices using descriptor system based realizations. For the computation of a proper $\mathds{C}_g$-stable RCF, no  particular properties of the underlying realizations need to be assumed. In particular, no $\mathds{C}_b$-stabilizability is required, because uncontrollable eigenvalues in $\mathds{C}_b$ are automatically removed. To determine factorizations with minimum degree denominator factors, the weakest requirement is that all controllable eigenvalues lying in $\mathds{C}_b$ are also observable. This requirement is fulfilled, for example,  if the underlying descriptor system realization is minimal, or only observable, or $\mathds{C}_b$-detectable.

For the computation of a stable RCF with inner denominator, the condition for the lack of poles in $\partial \mathds{C}_s$ (the boundary of the stability domain)  imposes the same condition on the eigenvalues of the pole pencil, provided the descriptor realization is minimal. However, the minimality condition on the descriptor system realization can be relaxed, by only requiring the lack of controllable eigenvalues in $\partial \mathds{C}_s$ (because the uncontrollable eigenvalues in $\partial \mathds{C}_s$ or  $\mathds{C}_b$ are automatically removed). The minimum degree requirement for $M(\lambda)$ imposes additionally the lack of unobservable eigenvalues in $\mathds{C}_s$.

In the light of the above considerations, both the \textbf{Procedure GRCF}, in conjunction with the pole selection and assignment schemes presented in Section \ref{sec:polsel}, as well as the \textbf{Procedure GRCFID}, in conjunction with the square-root-based computation of elementary inner factors described in Section \ref{sec:squareroot}, can be considered completely satisfactory numerical algorithms, which fulfill the standard requirements formulated in \cite{Mole78} for generality, numerical reliability and computational efficiency.


\begin{thebibliography}{10}
\expandafter\ifx\csname url\endcsname\relax
  \def\url#1{\texttt{#1}}\fi
\expandafter\ifx\csname urlprefix\endcsname\relax\def\urlprefix{URL }\fi
\expandafter\ifx\csname href\endcsname\relax
  \def\href#1#2{#2} \def\path#1{#1}\fi

\bibitem{Zhou96}
K.~Zhou, J.~C. Doyle, K.~Glover, Robust and Optimal Control, Prentice Hall,
  Upper Saddle River, 1996.

\bibitem{Varg17}
A.~Varga, Solving Fault Diagnosis Problems -- Linear Synthesis Techniques,
  Vol.~84 of Studies in Systems, Decision and Control, Springer International
  Publishing, 2017.
\newblock \href {http://dx.doi.org/10.1007/978-3-319-51559-5}
  {\path{doi:10.1007/978-3-319-51559-5}}.

\bibitem{Bele68}
V.~Belevitch, Classical Network Theory, Holden Day, San Francisco, 1968.

\bibitem{Vand77}
J.~Vandewalle, P.~Dewilde, On the irreducible cascade synthesis of a system
  with real rational transfer matrix, {IEEE} Trans. Circuits Syst. 24 (1977)
  481--494.
\newblock \href {http://dx.doi.org/10.1109/TCS.1977.1084377}
  {\path{doi:10.1109/TCS.1977.1084377}}.

\bibitem{Door90}
P.~{Van~Dooren}, Rational and polynomial matrix factorizations via recursive
  pole-zero cancellation, Linear Algebra Appl. 137/138 (1990) 663--697.
\newblock \href {http://dx.doi.org/10.1016/0024-3795(90)90144-2}
  {\path{doi:10.1016/0024-3795(90)90144-2}}.

\bibitem{Varg98a}
A.~Varga, Computation of coprime factorizations of rational matrices, Linear
  Algebra Appl. 271 (1998) 83--115.
\newblock \href {http://dx.doi.org/10.1016/S0024-3795(97)00256-5}
  {\path{doi:10.1016/S0024-3795(97)00256-5}}.

\bibitem{Oara99b}
C.~Oar\u{a}, A.~Varga, Minimal degree coprime factorization of rational
  matrices, SIAM J. Matrix Anal. Appl. 21 (1999) 245--278.
\newblock \href {http://dx.doi.org/10.1137/S0895479898339979}
  {\path{doi:10.1137/S0895479898339979}}.

\bibitem{Oara05}
C.~Oar\u{a}, Constructive solutions to spectral and inner-outer factorizations
  with respect to the disk, Automatica 41 (2005) 1855--1866.
\newblock \href {http://dx.doi.org/10.1016/j.automatica.2005.04.009}
  {\path{doi:10.1016/j.automatica.2005.04.009}}.

\bibitem{Verg79}
G.~Verghese, P.~{Van~Dooren}, T.~Kailath, Properties of the system matrix of a
  generalized state-space system, Int. J. Control 30 (1979) 235--243.
\newblock \href {http://dx.doi.org/10.1080/00207177908922771}
  {\path{doi:10.1080/00207177908922771}}.

\bibitem{Verg81}
G.~Verghese, B.~L\'evy, T.~Kailath, A generalized state-space for singular
  systems, IEEE Trans. Automat. Control 26 (1981) 811--831.
\newblock \href {http://dx.doi.org/10.1109/TAC.1981.1102763}
  {\path{doi:10.1109/TAC.1981.1102763}}.

\bibitem{Golu13}
G.~H. Golub, C.~F. {Van~Loan}, Matrix Computations, 4th Edition, John Hopkins
  University Press, Baltimore, 2013.

\bibitem{Oara97}
C.~Oar\u{a}, P.~V. Dooren, An improved algorithm for the computation of
  structural invariants of a system pencil and related geometric aspects, Syst.
  Control Lett. 30 (1997) 39--48.
\newblock \href {http://dx.doi.org/10.1016/S0167-6911(96)00078-3}
  {\path{doi:10.1016/S0167-6911(96)00078-3}}.

\bibitem{Door79a}
P.~{Van~Dooren}, The computation of {Kronecker's} canonical form of a singular
  pencil, Linear Algebra Appl. 27 (1979) 103--141.
\newblock \href {http://dx.doi.org/10.1016/0024-3795(79)90035-1}
  {\path{doi:10.1016/0024-3795(79)90035-1}}.

\bibitem{Varg95}
A.~Varga, On stabilization of descriptor systems, Syst. Control Lett. 24 (1995)
  133--138.
\newblock \href {http://dx.doi.org/10.1016/0167-6911(94)00017-P}
  {\path{doi:10.1016/0167-6911(94)00017-P}}.

\bibitem{Vidy11}
M.~Vidyasagar, Control System Synthesis: A Factorization Approach, "Morgan
  {\rm\&} Claypool", 2011.

\bibitem{Varg81b}
A.~Varga, A {Schur} method for pole assignment, IEEE Trans. Automat. Control 26
  (1981) 517--519.
\newblock \href {http://dx.doi.org/10.1109/TAC.1981.1102605}
  {\path{doi:10.1109/TAC.1981.1102605}}.

\bibitem{Hamm82}
S.~J. Hammarling, Numerical solution of the stable, non-negative definite
  {Lyapunov} equation, IMA J. Numer. Anal. 2 (1982) 303--323.
\newblock \href {http://dx.doi.org/10.1093/imanum/2.3.303}
  {\path{doi:10.1093/imanum/2.3.303}}.

\bibitem{Penz98a}
T.~Penzl, Numerical solution of generalized {Lyapunov} equations, Adv. Comput.
  Math. 8 (1998) 33--48.
\newblock \href {http://dx.doi.org/10.1023/A:1018979826766}
  {\path{doi:10.1023/A:1018979826766}}.

\bibitem{Varg17a}
A.~Varga, {DSTOOLS -- The Descriptor System Tools for MATLAB}, 2019,
  \url{https://sites.google.com/site/andreasvargacontact/home/software/dstools}.

\bibitem{Beel88}
T.~Beelen, P.~{Van~Dooren}, An improved algorithm for the computation of
  {Kronecker's} canonical form of a singular pencil, Linear Algebra Appl. 105
  (1988) 9--65.
\newblock \href {http://dx.doi.org/10.1016/0024-3795(88)90003-1}
  {\path{doi:10.1016/0024-3795(88)90003-1}}.

\bibitem{Benn99}
P.~Benner, V.~Mehrmann, V.~Sima, S.~{Van Huffel}, A.~Varga, {SLICOT} -- a
  subroutine library in systems and control theory, in: B.~N. Datta (Ed.),
  Applied and Computational Control, Signals and Circuits, Vol.~1,
  Birkh\"auser, 1999, pp. 499--539.
\newblock \href {http://dx.doi.org/10.1007/978-1-4612-0571-5_10}
  {\path{doi:10.1007/978-1-4612-0571-5_10}}.

\bibitem{Mole78}
C.~B. Moler, C.~F. {Van~Loan}, Nineteen dubious ways to compute the exponential
  of a matrix, SIAM Rev. 20 (1978) 801--836.
\newblock \href {http://dx.doi.org/10.1137/1020098}
  {\path{doi:10.1137/1020098}}.

\end{thebibliography}

\end{document}